\documentclass[twocolumn,showpacs,preprintnumbers,amsmath,amssymb]{revtex4}
\usepackage{amssymb}
\usepackage[dvips]{graphicx}
\begin{document}
\title{Isotope effect on the electron band structure of doped insulators}

\author{P.E. Kornilovitch $^{1}$ and A.S. Alexandrov$^{2}$}
 \affiliation{ $^1$Hewlett-Packard, MS 321A, 1000 NE Circle blvd,
Corvallis,
Oregon 97330, USA \\
$^2$Department of Physics, Loughborough University, Loughborough LE11\,3TU, UK\\
}

\begin{abstract}

Applying a continuous-time quantum Monte-Carlo algorithm we
calculate the exact coherent band dispersion and the density of
states of a two dimensional lattice polaron in the region of
parameters where any approximation might fail. We find an isotope
effect on the band structure, which is  different for different
wave-vectors of the Brillouin zone and depends on the radius and
strength of the electron-phonon interaction. An isotope effect on
the electron spectral function is also discussed.
\end{abstract}

 \vskip2pc]

\maketitle

The isotope substitution, where an ion mass $M$ is varied without
any change of electronic configurations, is a powerful tool in
testing the origin of electron correlations in solids. In
particular, a finite value of the isotope exponent $\alpha =-d\ln
T_{c}/d\ln M$ \cite{iso} proved that the superconducting phase
transition at $T=T_c$ is driven by the electron-phonon (e-ph)
interaction in conventional low-temperature superconductors.
Advances in the fabrication of isotope substituted samples made it
possible to measure a sizable isotope effect also in many
high-temperature superconductors. This led to a general conclusion
that phonons are relevant for high $T_{c}$. Essential features of
the isotope effect on $T_c$, in particular its large values in low
$T_{c}$ cuprates, an overall trend to decrease as $T_{c}$
increases, and a small or even negative $\alpha $ in some high
$T_{c}$ cuprates, could be understood in the framework of the
bipolaron theory of high-temperature superconductivity
\cite{aleiso}.

The most compelling evidence for (bi)polaronic carries in novel
superconductors was provided by the discovery of a substantial
isotope effect on the (super)carrier mass \cite{guo,kha}. The
effect was observed by measuring the magnetic field penetration
depth $\lambda_{H}$ of isotope-substituted copper oxides. The
carrier density is unchanged with the
isotope substitution of O$^{16}$ by O$^{18}$, so that the isotope effect on $%
\lambda_{H}$ measures directly the isotope effect on the carrier mass $%
m^{\ast }$. A carrier mass isotope exponent $\alpha_{m}=d\ln
m^{\ast }/d\ln M $ was observed, as predicted by the bipolaron
theory \cite{aleiso}. In ordinary metals, where the Migdal
adiabatic approximation is believed to be valid, $\alpha _{m}=0$
is expected. However, when the e-ph interaction is sufficiently
strong and electrons form polarons (quasiparticles dressed by
lattice distortions), their effective mass $m^{*}$ depends on $M$
as $m^{*}= m \exp (\gamma E_p/\omega)$. Here $m$ is the band mass
in the absence of the electron-phonon interaction, $E_p$ is the
polaron binding energy (polaron level shift), $\gamma$ is a
numerical constant less than $1$ that depends on the radius of the
electron-phonon interaction, and $\omega$ is a
characteristic phonon frequency (we use $\hbar=1$). In the expression for $%
m^*$, only the phonon frequency depends on the ion mass. Thus
there is a
large isotope effect on the carrier mass in (bi)polaronic conductors, $%
\alpha_{m} = (1/2)\ln (m^*/m)$ \cite{aleiso}, in contrast with the
zero isotope effect in ordinary metals.

Recent high resolution angle resolved photoemission spectroscopy
(ARPES) \cite{lan0} provided another compelling evidence for a
strong e-ph interaction in the cuprates. It revealed a fine phonon
structure in the electron self-energy of the underdoped
La$_{2-x}$Sr$_x$CuO$_4$ samples \cite {shencon}, and a complicated
isotope effect on the electron spectral function in Bi2212
\cite{lan}. Polaronic carriers were also observed in colossal
magneto-resistance manganites including their low-temperature
ferromagnetic phase, where an isotope effect on the residual
resistivity was measured and explained \cite{alezhao}.

These and many other experimental and theoretical observations
point towards unusual e-ph interactions in complex oxides, which
remain to be quantitatively addressed. While the many-particle
e-ph problem has been solved in the weak-coupling, $\lambda\equiv
E_p/zt \ll 1$ \cite{mig,shri,eli}, and in the strong-coupling
$\lambda \gg 1$ \cite{ale} limits, any analytical or even
semi-analytical approximation (i.e. dynamic mean-field approach in
finite dimensions) is refutable in the relevant intermediate
region of the coupling strength, $\lambda \simeq 1$, and of the
adiabatic ratio, $\omega/t \simeq 1$. Here $t$ and $z$ are the
nearest neighbor hopping integral and the coordination number of
the rigid lattice, respectively.

Advanced variational \cite{tru}, direct diagonalization
\cite{alekab}, and quantum Monte Carlo (QMC) \cite{hr,lag,prok}
techniques addressed the problem in the intermediate region of
parameters, but mainly in the framework of the Holstein model
\cite{hol} with a local (short-range) e-ph interaction, or in the
continuous (effective mass) approximation for the bare electron
band \cite{prok}. However, in cuprates there is virtually no
screening of $c-$axis polarised optical phonons because an upper
limit for the out-of-plane plasmon frequency ($\lesssim 200$
cm$^{-1}$\cite{marel}$)$
is well below the characteristic phonon frequency $\omega \simeq 1000$ cm$%
^{-1}$. Hence, the unscreened long-range Fr\"{o}hlich interaction
is  important in cuprates and other ionic charge-transfer
insulators \cite{ale2,alekor,feh,bon,tru}. A parameter-free
estimate \cite{alebra} yields $E_{p}\simeq 1$ eV with this
interaction alone, which is larger than a magnetic (i.e. exchange
$J$) interaction almost by one order of magnitude. Qualitatively,
a longer-range el-ph interaction results in a lighter mass of
dressed carries ($\gamma <1$) because the extended lattice
deformation changes gradually in the process of tunnelling through
the lattice \cite{ale2,alekor}. Also in the intermediate and
strong-coupling regime the finite bandwidth is important, so that
the effective mass approximation cannot be applied.

In this Letter we calculate isotope effects on the whole coherent
band-structure and DOS of a two-dimensional lattice polaron with
short- and long-range e-ph interactions by applying a
continuous-time QMC algorithm. Unlike the strong-coupling limit
\cite{aleiso} the isotope effect depends on the wave vector in the
intermediate region of parameters. We also discuss the isotope
effect on the electron spectral function including its incoherent
part.

Our model Hamiltonian describes any-radius e-ph interaction of an
electron (or hole) on a square-lattice plane with linearly
polarized vibrations of all ions of the 3D crystal \cite{alekor},
\begin{equation}
H=-\sum_{\mathbf{m,n }}[t(\mathbf{m-n}) c^{\dagger}_{\mathbf{m}}c_{\mathbf{n}%
} + f(\mathbf{m-n})\xi_{\mathbf{n}}
c^{\dagger}_{\mathbf{m}}c_{\mathbf{m}}] + \omega
\sum_{\mathbf{q}}b^{\dagger}_{\mathbf{q}} b_{\mathbf{q}}.
\label{eq:one}
\end{equation}
Here $c_{\mathbf{m}}$, and $b_{\mathbf{q}}$ are annihilation
electron and phonon operators, respectively, $t(\mathbf{m-n})$ is
the hopping integral,
which is nonzero only for the nearest neighbors, $t(\mathbf{a}) \equiv t$ ($%
\mathbf{a}$ is the primitive lattice vector),
$\xi_{\mathbf{n}}=(2NM\omega
)^{-1/2}\sum_{\mathbf{q}}e^{i\mathbf{q\cdot
n}}b_{\mathbf{q}}+H.c.$ is the
displacement operator at cite $\mathbf{n}$, $N$ is the number of cells, and $%
f(\mathbf{m-n})$ is a ``force'' applied to the electron at site
$\mathbf{m}$ due to the ion displacement at cite $\mathbf{n}$.

In the strong-coupling, $\lambda \gg 1$, and non-adiabatic limit
$\omega \gtrsim t $ one can apply the Lang-Firsov transformation
\cite{fir}
to obtain the ground state energy $E_{p}=(2M\omega) ^{-2}\sum_{\mathbf{m}}f^{2}(%
\mathbf{m})$ and the coherent polaron band dispersion $\epsilon_{\mathbf{k}%
}= E_{\mathbf{k}} \exp(-g^2)$ with the polaron narrowing exponent $%
g^{2}=(2M\omega ^{3})^{-1}\sum_{\mathbf{m}}\left[ f^{2}(\mathbf{m})-f(%
\mathbf{m})f(\mathbf{m+a})\right]$ and the ``bare'' band dispersion $E_{%
\mathbf{k}}$. The band-structure isotope exponent $\alpha_b$ does
not depend on the wave-vector $\mathbf{k}$ in this limit, because
e-ph interactions do not change the band topology,
\begin{equation}
\alpha_b \equiv - \frac{\partial \ln
\epsilon_{\mathbf{k}}}{\partial \ln M} = \frac{g^2}{2}.
\label{eq:two}
\end{equation}
It is the same as $\alpha_m$. If the interaction is short-range, $f(\mathbf{%
m-n})\propto \delta _{\mathbf{m, n}}$ (the Holstein model), then $%
g^{2}=E_{p}/\omega $. Generally, one has $g^{2} = \gamma
E_{p}/\omega $ with the numerical coefficient
$\gamma =(1-\sum_{\mathbf{m}}f(\mathbf{m})f(\mathbf{m+a}))/\sum_{\mathbf{%
n}}f^{2}(\mathbf{n})$. $\gamma$ can be significantly less than $1$
for a long-range (Fr\"ohlich) interaction \cite{ale2}, such as the
unscreened interaction with $c$-axis vibrations of apex oxygen
ions in the cuprates \cite{alekor}:
\begin{equation}
f(\mathbf{m-n})\propto(|\mathbf{m}-\mathbf{n}|^{2}+1)^{-3/2}.
\label{eq:four}
\end{equation}
Here the distance between the in-plane and apex cites $\sqrt{|\mathbf{m}-%
\mathbf{n}|^2+1}$ is measured in units of the in-plane lattice
constant $a$, and the apex-plane distance is taken to be also $a$.
Thus the strong-coupling isotope exponent, Eq.~(\ref{eq:two})
turns out to be numerically smaller for a longer-range e-ph
interaction compared with a short-range interaction of the same
strength $E_p$.

\begin{figure}[tbp]
\begin{center}
\includegraphics[angle=-00,width=0.40\textwidth]{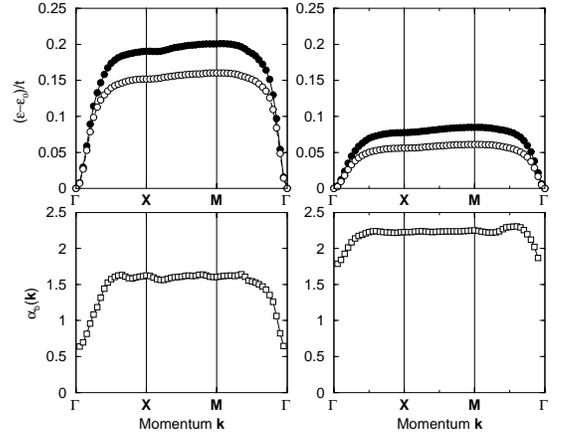}
\end{center}
\caption{Top panels: small Holstein polaron band dispersions along
the main
directions of the two-dimensional Brillouin zone. Left: $\protect\lambda %
=1.1 $, right: $\protect\lambda =1.2$. Filled symbols is $\protect\omega %
=0.70\,t$, open symbols is $\protect\omega =0.66\,t$. Lower
panels: the band
structure isotope exponent for $\protect\lambda =1.1$ (left) and $\protect%
\lambda =1.2$ (right).}
\end{figure}

QMC results (see below) show that this tendency also holds in the
intermediate coupling regime, but the isotope exponent becomes a
nontrivial function of the wave vector: $\alpha _{b} = \alpha
_{b}(\mathbf{k)}$, because e-ph interactions substantially modify
the band topology in this regime. We apply a continuous-time
path-integral QMC algorithm developed by one of us \cite {kor},
which is free from any systematic finite-size and finite-time-step
errors. The finite-temperature errors are exponentially small when
the simulation temperature is smaller than the phonon frequency.
The method allows for \emph{exact} calculation of the whole
polaron band dispersion on any-dimensional infinite lattice with
any-range e-ph interaction using the many-body path integral,
\begin{equation}
\epsilon _{0}=\frac{\int
\mathcal{D}\mathbf{r}\mathcal{D}\mathbf{\xi } \left[ - \frac{\partial w}{\partial \beta }\right] }{%
\int \mathcal{D}\mathbf{r}\mathcal{D}\mathbf{\xi }\cdot w},
\label{eq:five}
\end{equation}
\begin{equation}
\epsilon _{\mathbf{k}}-\epsilon _{0}=-\frac{1}{\beta }\ln \left\{
\frac{\int
\mathcal{D}\mathbf{r}\mathcal{D}\mathbf{\xi }\cdot w\cdot e^{i\mathbf{k}%
\cdot \Delta \mathbf{r}}}{\int \mathcal{D}\mathbf{r}\mathcal{D}\mathbf{\xi }\cdot w%
}\right\} ,  \label{eq:six}
\end{equation}
whre $\epsilon _{0}$ is the ground state energy. Configurations
described by the electron position vector $\mathbf{r}$ and ion
displacements $\mathbf{\xi }$, are obtained at imaginary time
$\tau =\beta =1/(k_{B}T)$ from the configurations at $\tau =0$ by
shifting along the lattice by vectors $\Delta \mathbf{r}$. $\int
\mathcal{{D}\mathbf{r}{D}\mathbf{\xi }}$ integrates over
single-electron paths with all possible shifts satisfying the
twisted boundary conditions with the weight $w(\mathbf{r,\xi })$.
The statistics for any number of $\mathbf{k}$ points in the
Brillouin zone are collected during a single QMC run \cite{rev}.

\begin{figure}[tbp]
\begin{center}
\includegraphics[angle=-00,width=0.40\textwidth]{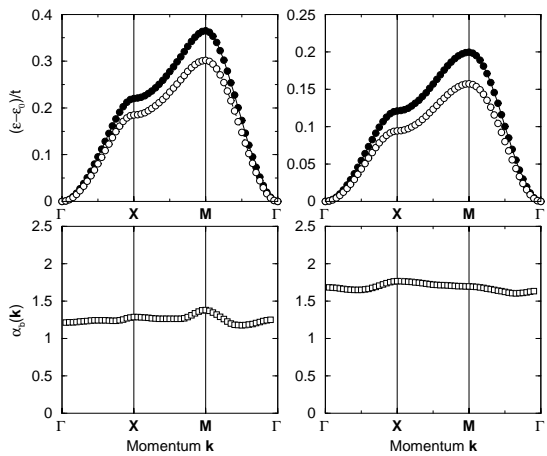}
\end{center}
\caption{Top panels: small Fr\"ohlich polaron band dispersions
along the
main directions of the two-dimensional Brillouin zone. Left: $\protect%
\lambda = 2.5$, right: $\protect\lambda = 3.0$. Filled symbols is $\protect%
\omega = 0.70 \, t$, open symbols is $\protect\omega = 0.66 \, t$.
Lower panels: the band structure isotope exponent for
$\protect\lambda = 2.5$ (left) and $\protect\lambda = 3.0$
(right).}
\end{figure}

The isotope exponent $\alpha _{b}(\mathbf{k})$ and DOS are
presented in Figs.~(1,2), and Fig.3, respectively, for the small
Holstein polaron (SHP) with the short-range interaction, and for
the small Fr\"{o}hlich polaron (SFP) with the force given by
Eq.~(\ref{eq:four}). The polaron spectra are calculated for two
phonon frequencies, $\omega =0.70\,t$ and $\omega
=0.66\,t $, whose difference corresponds to a substitution of O$^{16}$ by O$%
^{18}$ in cuprates. There is a significant change in the
dispersion law (topology) of SHP, Fig.1, which is less significant
for SFP, Fig.2, rather than a simple band-narrowing (see also
\cite{kor}). As a result, the isotope exponent
\begin{equation}
\alpha _{b}(\mathbf{k})\approx 8\frac{\epsilon _{\mathbf{k}}^{16}-\epsilon _{%
\mathbf{k}}^{18}}{{\epsilon _{\mathbf{k}}^{16}}},
\label{eq:seven}
\end{equation}
is $\mathbf{k}$-dependent, Fig.1 and Fig.2 (lower panels). The strongest dispersion of $%
\alpha _{b}$ is observed for SHP. Importantly, the isotope effect
is suppressed near the band edge in contrast with the
$\mathbf{k}$-independent strong-coupling isotope effect,
Eq.~(\ref{eq:two}). It is less dispersive for SFP, especially at
larger $\lambda$, where the polaron energy band is well described
by the strong-coupling Lang-Firsov expression.

Polaron DOS, $N(E) \equiv \sum_{\mathbf{k}} \delta(E-\epsilon_{\mathbf{k}})$%
, Fig.3, is less sensitive to topology, approximately scaling with
the renormalised bandwidth. It reveals a giant isotope effect near
the van Hove singularity because the bandwidth changes
significantly with $\omega$ even in the intermediate-coupling
regime.

The coherent motion of small polarons leads to metallic conduction
at low temperatures. Our results, Figs.1-3, show that there should
be anomalous isotope effects on low-frequency kinetics and
thermodynamics of polaronic conductors which  depend on the
position of the Fermi level in the polaron band. In fact, such
effects have been observed in ferromagnetic oxides at low
temperatures \cite{alezhao}, and in cuprates \cite{guo,kha}. To
address ARPES isotope exponents \cite{lan} one has to calculate
the electron spectral function $A(\mathbf{k}, E)$ taking into
account phonon side-bands (i.e. off diagonal transitions) along
with the coherent polaron motion (diagonal transitions). Using the
$1/\lambda$ expansion one obtains \cite {ale}
\begin{equation}
A(\mathbf{k},E )=\sum_{l=0}^{\infty } \left[ A_{l}^{(-)}(\mathbf{k}%
,E)+A_{l}^{(+)}(\mathbf{k},E )\right] ,  \label{eq:eight}
\end{equation}
where
\begin{eqnarray}
&& A_{l}^{(-)}(\mathbf{k},E ) = \frac{Z\left[ 1-n(E-l\omega) \right]}{%
(2N)^{l}l!} \cr &\times& \sum_{\mathbf{q}_{1},...\mathbf{q}%
_{l}}\prod_{r=1}^{l}|\gamma (\mathbf{q}_{r})|^{2} \delta [ E -
l\omega - \zeta (\mathbf{k}_l^-)] ,  \nonumber  \label{eq:nine}
\end{eqnarray}
and
\begin{equation}
A_{l}^{(+)}(\mathbf{k},\omega ) = \frac{Z \cdot
n(E+l\omega)}{(2N)^{l}l!}
\sum_{\mathbf{q}_{1},...\mathbf{q}_{l}} \prod_{r=1}^{l}| \gamma (\mathbf{q}%
_{r})|^{2} \delta [ E - l\omega - \zeta (\mathbf{k}_l^+)].
\label{eq:ten}
\end{equation}
Here $Z=\exp(-E_p/\omega)$, $\zeta(\mathbf{k})=\epsilon_{\mathbf{k}}-\mu$, $%
\mu$ is the chemical potential, $n(E)=[\exp (\beta E)+1]^{-1}$, $\mathbf{k}%
_l^\pm=\mathbf{k}\pm \sum_{r=1}^{l} \mathbf{q}_r$ and
$\gamma(\mathbf{q})$
is the Fourier transform of the force $f(\mathbf{m})=N^{-1}M^{1/2}%
\omega^{3/2}\sum_{\mathbf{q}}\gamma (\mathbf{q})e^{i\mathbf{q\cdot
m}}$.

\begin{figure}[tbp]
\begin{center}
\includegraphics[angle=-00,width=0.40\textwidth]{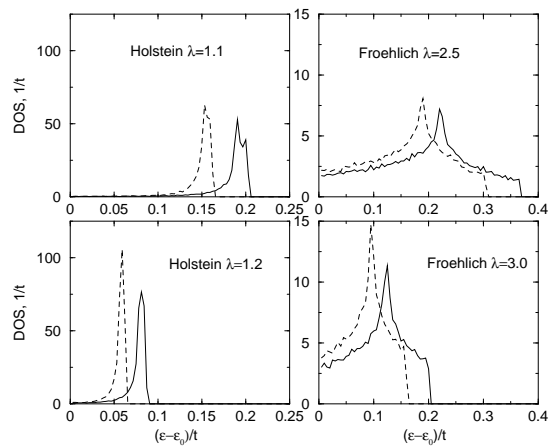}
\end{center}
\caption{Isotope effect on the polaron density of states. Solid line is $%
\protect\omega = 0.70 \, t.$, dashed line is $\protect\omega =
0.66 \, t$.}
\end{figure}

Clearly, Eq.~(\ref{eq:eight}) is in the form of a perturbative
multi-phonon expansion. Each contribution $A_{l}^{(\pm
)}(\mathbf{k},E )$ to the spectral function describes the
transition from the initial state $\mathbf{k}$ of the polaron band
to the final state $\mathbf{k}_l^\mp$ with the emission (or
absorption) of $l$ phonons. Different from the conventional Migdal
self-energy \cite{mig} the electron spectral function comprises
two different parts in the strong-coupling limit. The first
($l=0$) $\mathbf{k}$-dependent term arises from the coherent
polaron tunnelling, $A_{coh}(\mathbf{k},E )= \left[
A_{0}^{(-)}(\mathbf{k},E ) + A_{0}^{(+)}(\mathbf{k},E ) \right] =
 Z \delta(E -\zeta_{\mathbf{k}})$ with a suppressed spectral
weight $Z \ll 1$. The second \textit{incoherent} part
$A_{incoh}(\mathbf{k},E )$ comprises all terms with $l \geq 1$. It
describes excitations accompanied by emission and absorption of
phonons. We note that its spectral density spreads over a wide
energy range of about twice the polaron level shift $E_{p}$, which
might be larger than the unrenormalised bandwidth $2zt$ in the
rigid lattice. On the contrary, the coherent part shows a
dispersion only in the energy window of the order of the polaron
bandwidth. It is important that the $incoherent$
background $A_{incoh}(\mathbf{k},E )$ is dispersive (i.e. $\mathbf{k}$%
-dependent) for the long-range interaction. Only in the Holstein
model with the short-range dispersionless e-ph interaction
$\gamma(\mathbf{q}) = const$ the incoherent part has no
dispersion.

Using Eq.~(\ref{eq:eight}) one readily predicts an isotope effect
on the coherent part dispersion $\epsilon_{\mathbf{k}}$ and its
spectral weight $Z$, \emph{and}
also on the incoherent background because $Z$, $\gamma(\mathbf{q})$, and $%
\omega$ all depend on $M$. While our prediction is qualitatively
robust it is difficult to quantify the ARPES isotope effect in the
intermediate region of parameters. The spectral function,
Eq.~(\ref{eq:eight}), is applied in
the strong-coupling limit $\lambda \gg1$. While the main sum rule $%
\int_{-\infty }^{\infty }dE A(\mathbf{k},E )=1$ is satisfied, the
higher-momentum integrals $\int_{-\infty }^{\infty }dE E^{p}
A(\mathbf{k},E ) $ with $p>0$, calculated with
Eq.~(\ref{eq:eight}) differ from their exact values \cite{kor2} by
an amount proportional to $1/\lambda$. The difference is due to
partial ``undressing'' of high-energy excitations in the
side-bands, which is beyond the lowest order $1/\lambda$
expansion. The role of electronic correlations should be also
addressed in connection with ARPES. While the results shown in
Figs. 1-3 describe band-structure isotope effects in
slightly-doped conventional and Mott-Hubbard insulators with a few
carriers, their spectral properties could be significantly
modified by the polaron-polaron interactions \cite{dev}, including
the bipolaron formation \cite{aledent} at finite doping. On the
experimental side, separation of the coherent and incoherent parts
in ARPES remains rather controversial \cite{cam}.

In conclusion, we have calculated the isotope effect on the band
structure of doped insulators employing a continuous-time quantum
Monte-Carlo algorithm, and  found its dispersion in the
intermediate coupling regime which essentially depends on the
radius and strength of the electron-phonon interaction. Using the
strong-coupling electron spectral function we predicted an isotope
effect on the weight and dispersion of the coherent part, and on
the incoherent background. The exact isotope exponents, Fig. 1 and
Fig.2, could be instrumental in understanding of current and
future experiments with isotope substituted oxides, and in
assessing different analytical and semi-analytical approximations.

We would like to thank J. Devreese, J. Hirsch, A. Lanzara, J.
Samson, and S. Trugman for illuminating discussions. ASA
acknowledges support of this work by the Leverhulme Trust
(London).

\end{document}